# Enhanced UV Light emission from Silicon nanoparticles induced by Au ion implantation

**Akhilesh Singh[1], Karol G. Gryczynski[1], Bibhu Rout[1], Jianyou Li[1], Floyd McDaniel[1], Arup Neogi[1], Gayatri Sahu[2], Durga P. Mahapatra[2]**

**[1]University of North Texas, TX, USA, [2]Institute of Physics, Orissa, India**

E-mail: arup@unt.edu
WWW: http://www.phys.unt.edu/research/photonic/

Although silicon is the preeminent semiconductor for electronics, due to the indirect nature of bandgap in bulk system, it is currently not possible to use it for optoelectronics or photonics application. By reducing the dimension of 3D bulk Silicon using 1D nanowires or quantum dots it may be possible to achieve direct bandgap transition for photonics. This can lead to the possibility of achieving light emission from silicon nanostructures resulting in industrially viable monolithically integrated Si based optoelectronic devices and systems. Current techniques based on chemical etching processes or ion-implantation in $SiO_2$ has resulted in weak light emission[i ii iii] . This is because the decay of carriers is predominantly non-radiative due to Auger recombination or inter-valley scattering. In this work we demonstrate a novel technique for fabricating silicon with light emitting properties using gold nanoparticles.

In this work light emission from Si(100) wafers was achieved using sequential ion beam treatments meant to create nano-particles within the wafers. Silicon quantum dots are formed due to the recrystallization of Si nanoparticles in pre-amorphized Si layer in the absence of oxygen. Recrysallization of Si nanoparticles with gold ion was found to enhance the light emission in the ultraviolet wavelength range. Furnace annealing of the Si substrate following the gold implantation process result in a migration of Si nanoparticles towards the surface and yields in enhanced light emission. Temperature and time-resolved photoluminescence (PL) measurements have been performed to study to the nature of optical emission from the Si nanoparticles formed by Au ion implantation.

Temperature dependent PL measurements were performed using the 325 nm excitation which results in PL emission peaks from Si nanoparticles at ~ 3.28 eV at room temperature [Fig. 1]. These peaks split at lower temperature [Fig. 2]; this is most likely due to a distribution of different size nano-particles. It is also interesting to note that oven annealing increases the brightness of emission by about two orders in magnitude.

Time resolved photoluminescence measurements at room temperature [Fig 3.] show a long lifetime with multiple components. At low temperatures (27 K) [Fig. 4], the PL lifetimes at the various exciton emission peaks are significantly different. The lifetime of the emission from higher energy states (3.35 eV) is in the sub-nanosecond regime indicating direct bandgap emission from Si quantum confined structures. The lifetime of emission from the low energy peak is over a few tens of ns.

Preliminary results indicate that emissions originate from the Si quantum dots. The emission energy from Si quantum dots is not influenced by Au ion implantation. The enhancement in PL is likely due to passivation of Si dots or non-resonant plasmonic affects due to the presence of gold ions increases the luminescence intensity.

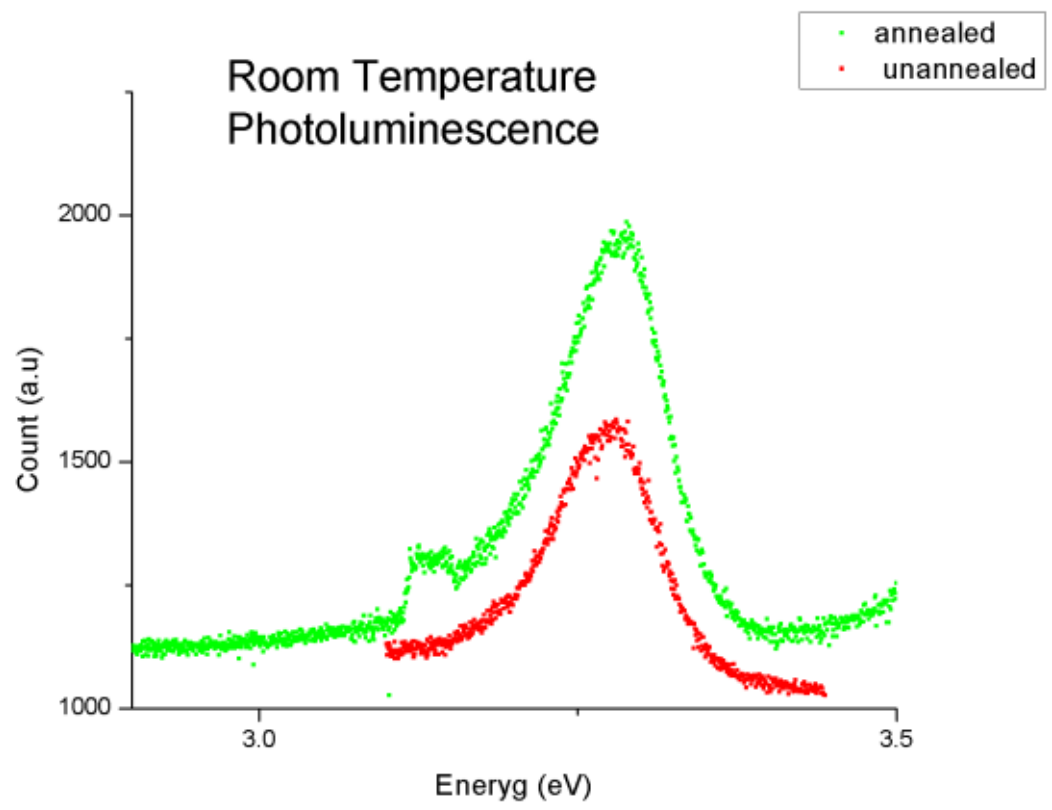

Figure 1: PL of Au ion implanted Si wafers at 300 K

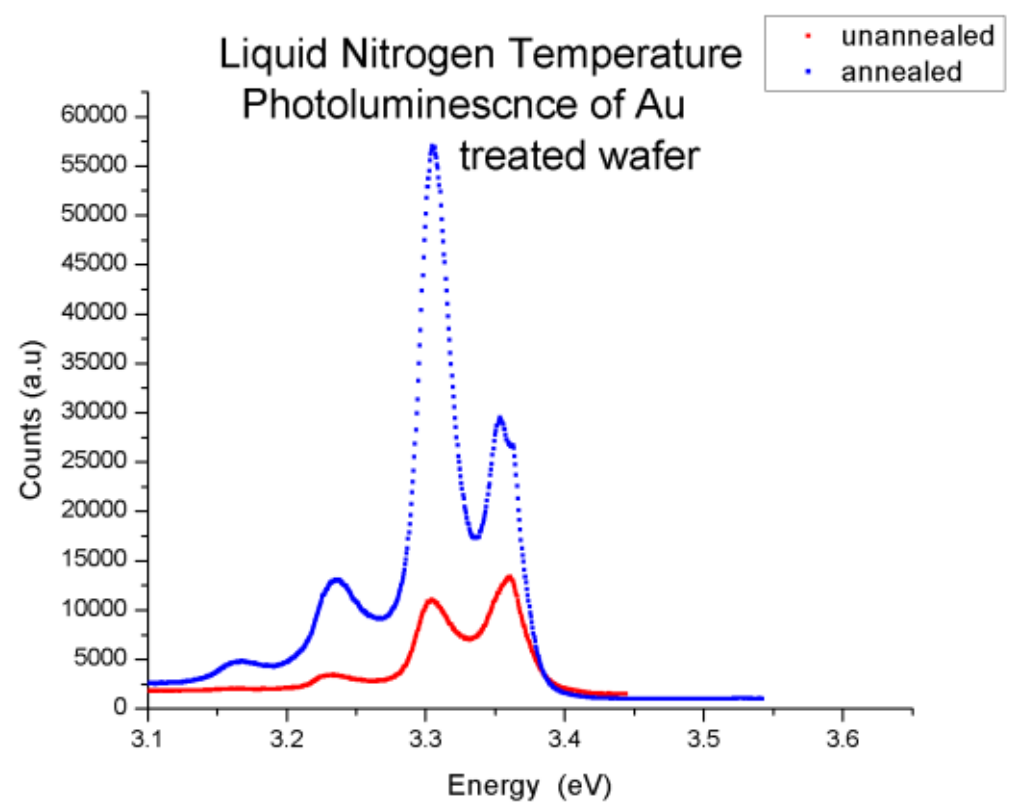

Figure 1: PL of Au ion implanted Si wafers at 77 K

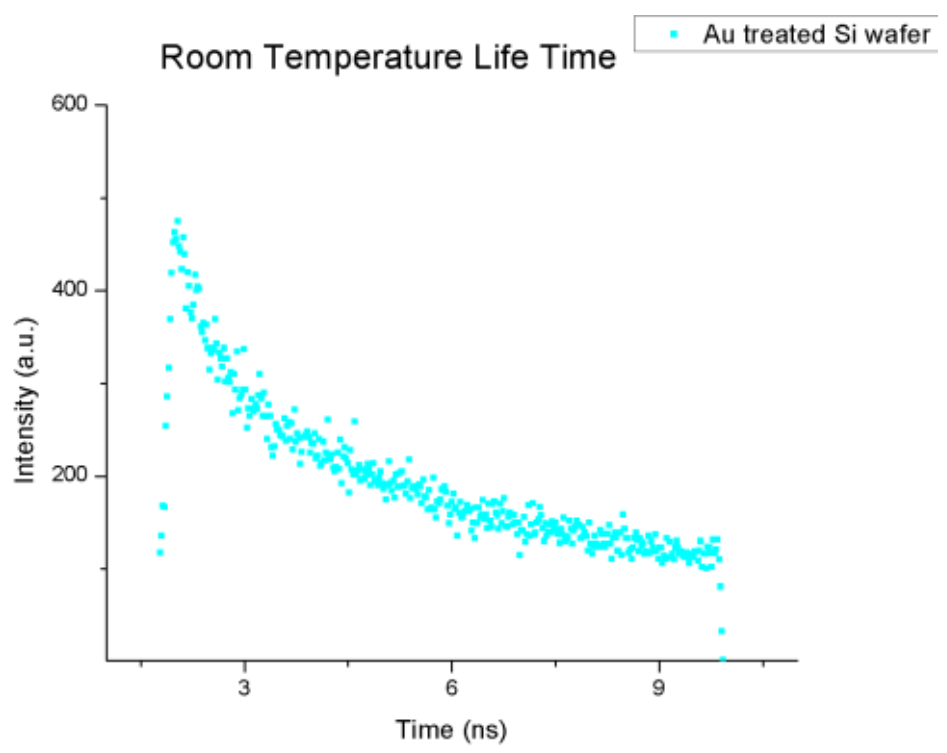

Figure 3: PL Lifetime of Au ion implanted Si nanoparticles at 300 K

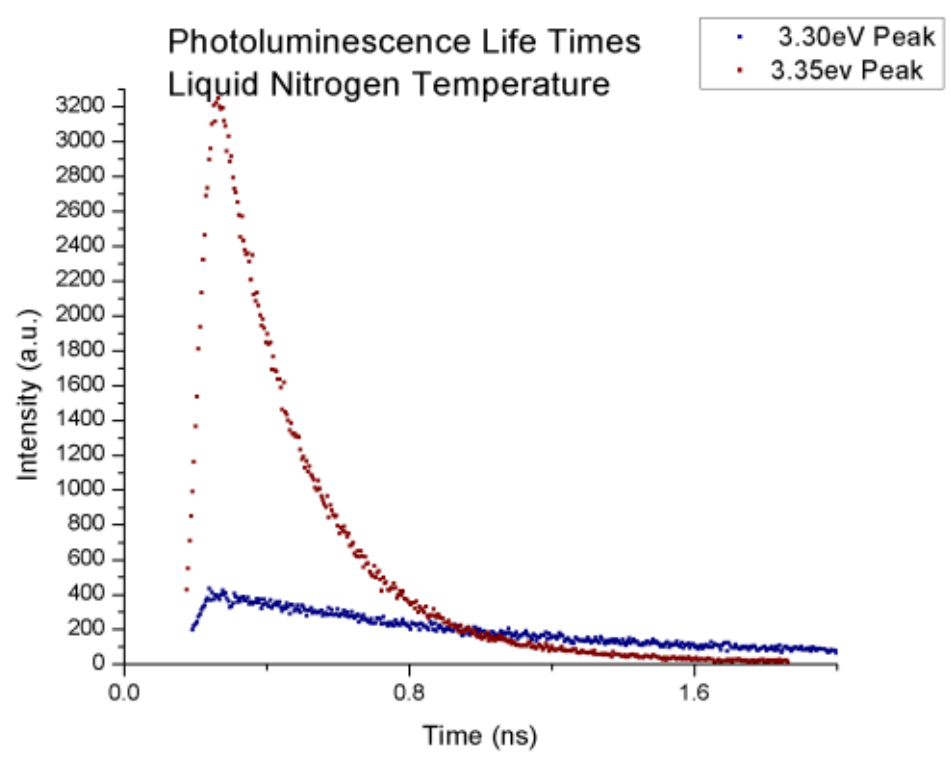

Figure 4: PL Lifetime of individual peaks of annealed Au treated Si wafer

This work is supported by National Science Foundation and Welch Foundation

---